\documentclass[11pt,preprint]{aastex}

\usepackage{emulateapj5}
\usepackage{epsf}
\usepackage{graphics}

\newcommand{\kms}{km\,s$^{-1}$}

\newenvironment{inlinefigure}{
\def\@captype{figure}
\noindent\begin{minipage}{0.999\linewidth}\begin{center}}
{\end{center}\end{minipage}\smallskip}

\shorttitle{HI kinematics in a massive spiral galaxy at z=0.89}
\shortauthors{Koopmans \& de Bruyn}

\begin{document}

\title{HI kinematics in a massive spiral galaxy at z=0.89}
\author{L.V.E. Koopmans} \affil{Space Telescope Science Institute,
3700 San Martin Drive, Baltimore, MD 21218, USA}

\smallskip

\author{A.G. de Bruyn} \affil{NFRA--ASTRON, P.O. Box 2, Dwingeloo 9700
AA, The Netherlands\\Kapteyn Astronomical Institute P.O. Box 800,
NL-9700 AV Groningen, Netherlands}

\smallskip

\begin{abstract}
We present a kinematic model of the neutral-hydrogen in the spiral
galaxy of the lens system PKS1830$-$211, based on a Multi Element
Radio-Linked Interferometer (MERLIN) 1.4--GHz radio map and the
integrated and redshifted 21--cm hydrogen absorption-line profile as
measured with the Westerbork Synthesis Radio Telescope (WSRT).
Degeneracies in the models do not allow a unique determination of the
kinematic center and forthcoming deeper Hubble Space Telescope
observations with the Advanced Camera for Surveys (ACS) are required
to break this degeneracy. Even so, we measure the inclination of the
hydrogen disk: $i=(17-32)^\circ$, indicating a close-to face-on spiral
galaxy.  The optical depth increases with radius over the extent of
the Einstein ring, suggesting HI depletion towards the lens
center. The latter could be due to star formation or conversion of HI
in to molecular hydrogen because of a higher metalicity/dust content
in the galaxy center. The neutral hydrogen optical depth gives $N_{\rm
HI}=2 \times 10^{21}$~cm$^{-2}$ at $r=5.0 \,h^{-1}_{70}$~kpc in the
disk ($T_{\rm s}=100$~K), comparable to local spiral galaxies.  Our
study shows that planned new radio telescopes (i.e.~ALMA, LOFAR and
SKA) -- which will discover large numbers of similar lens systems --
are powerful new tools to probe the internal structure, kinematics and
evolution of spiral galaxies in detail to $z\ga 1$, as well as their
neutral-hydrogen (and molecular) content, thereby complementing
studies of HI emission from spirals at $z \la 1$ and damped
Ly--$\alpha$ systems to $z\gg 1$.
\end{abstract}

\keywords{gravitational lensing --- galaxies: structure --- ISM:
kinematics and dynamics ---  line: profiles}

\section{Introduction}

The gravitational lens PKS 1830--211 (e.g. Pramesh Rao \& Subrahmanyan
1988; Subrahmanyan et al. 1990) is the brightest known radio lens in
the sky. The source at $z$=2.51 (Lidman et al. 1999) is extended and
lensed into an Einstein ring (e.g. Jauncey et al. 1991) by a
high-redshift spiral galaxy at $z$ = 0.89 (Wiklind \& Combes 1996;
Gerin et al. 1997; Mathur \& Nair 1997; Chengalur, de Bruyn \&
Narasimha 1999). In addition, a time-delay of 26$\pm$5 days between
the two images of the lensed core has been measured (Lovell et
al. 1998; see also Wiklind \& Alloin 2002). New monitoring data in
hand is expected to yield an improved measurement of this delay
(J.\,Winn, private communications). The structure of the Einstein ring
can be used to place constraints on the mass distribution of the lens
(Kochanek \& Narayan 1992; Nair, Narasimha, \& Rao 1993; Wiklind \&
Alloin 2002).  Were it not that PKS 1830--211 is seen through the
Galactic disk (i.e. $l=+12^\circ$, $b=-5^\circ$) -- leading to
considerable optical extinction ($A_I=0.8$ mag) and a field crowed by
stars (e.g. Winn et al. 2002; Courbin et al. 2002) -- this lens system
would be well-suited to determine the Hubble Constant.  Due to the
severe extinction it has proven difficult to precisely determine the
center of the lens galaxy. A number of mutually inconsistent lens
centers have been published over the years (e.g. Kochanek \& Narayan
1992; Nair et al. 1993; Lehar et al. 2000). Most recently, using the
same HST observations, two different interpretations of the lens
center were published (Winn et al. 2002; Courbin et al. 2002). The
inability to measure the lens center with the currently available
observations severely limits the use of PKS\,1830$-$211 to determine
H$_0$. Forthcoming HST--ACS observations, however, will be able to
settle this question.

Besides being interesting because of H$_0$, the lens galaxy in
PKS\,1830--211 is the highest-redshift spiral lens galaxy known. The
brightness of the lensed source has allowed a unique study of the
molecular (e.g. CO and OH) and atomic (i.e. HI) interstellar medium
(ISM) content (Wiklind \& Combes 1996; Gerin et al. 1997; Mathur \&
Nair 1997; Chengalur, de Bruyn \& Narasimha 1999) in a spiral galaxy
as seen $\approx$8\,$h_{70}^{-1}$~Gyr ago. Although only a single galaxy,
it provides a test case for similar systems anticipated to be
discovered with planned instruments such as ALMA, LOFAR and SKA.

The time base-line covered by these galaxies provide enough leverage
to study the evolution of the mass-distribution and ISM of spiral
galaxies in the redshift-regime where most of their evolution is
expected to occur and also connects up to studies of damped
Ly--$\alpha$ systems (e.g. Kauffmann 1996; Mo, Mao \& White 1998).

In this letter, we make a first attempt to model the kinematics of HI
in this high-redshift spiral galaxy, improving upon a previous more
simple analysis by Chengalur et al. (1999). One of the initial aims,
i.e. the determination of an independent kinematic lens center, has
proven too difficult given the current data, but constraints on the
radial distribution of HI, its column density and kinematics are
obtained.

\begin{inlinefigure}
\vspace{0.3cm}
\begin{center}
\resizebox{0.95\textwidth}{!}{\includegraphics{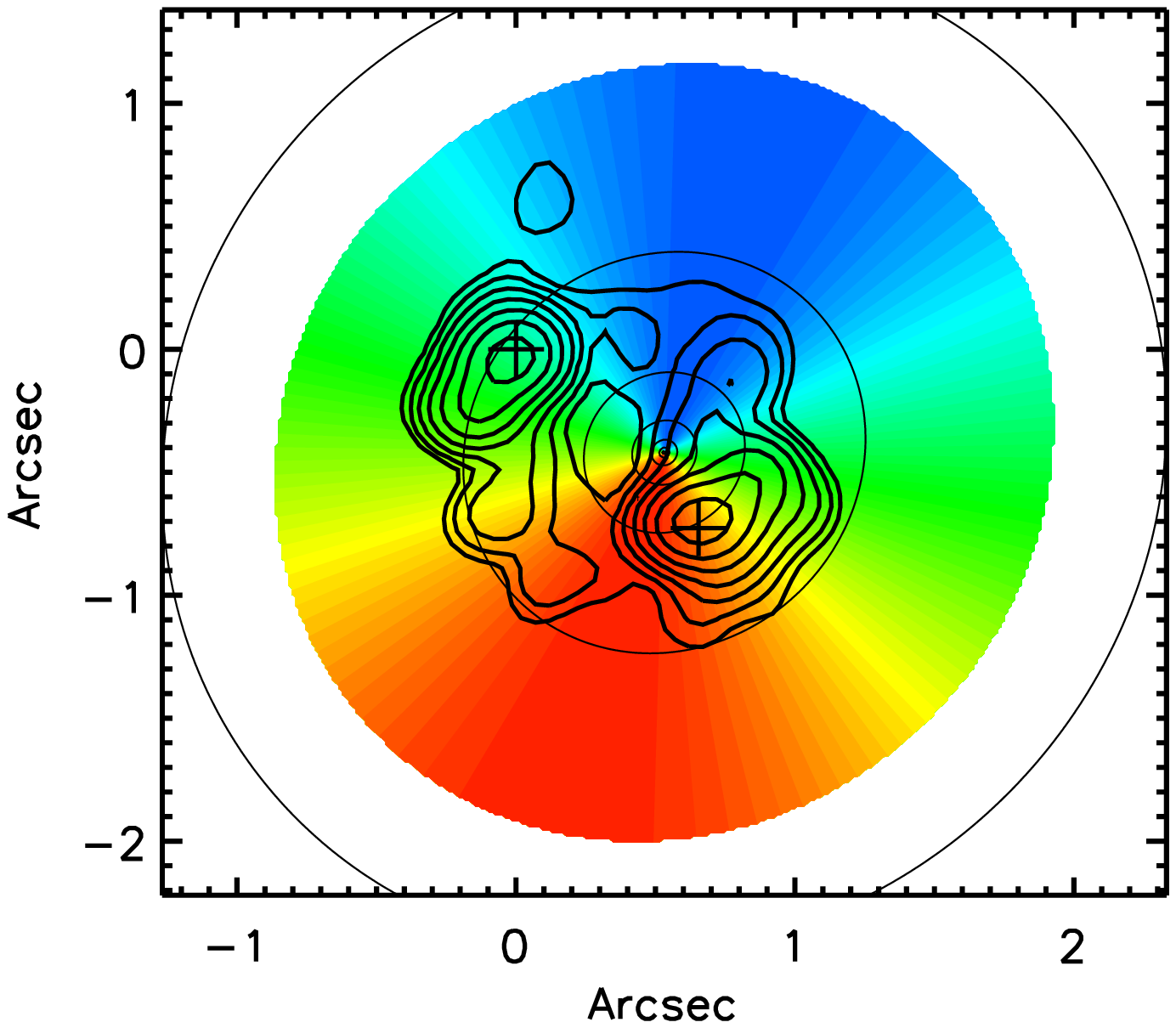}}
\end{center}
\figcaption{The MERLIN 1.4--GHz continuum maps of PKS\,1830--211
(North is up and East is left). Contours start at 25 mJy/beam and
increase by factors of two. Overplotted is the best-fit kinematic
model for the lens center of Nair et al. (1993). The two crosses
indate the lensed cores A (NE) and B (SW). The thin elliptical
contours indicate constant optical depth increasing by factors of 2
outward. The optical depth contour passing through image B has
$\tau$=0.064.}
\end{inlinefigure}

\section{MERLIN \& WSRT Observations}

MERLIN long-track observations of PKS1830--211 were obtained at 1.4,
1.7 and 4.9 GHz on, respectively, March 26, 25 and 8 in 1999. The data
were flux and phase calibrated at the telescope. We subsequently map
the data using DIFMAP (Shepherd 1997) following standard procedures,
but perform additional self-calibration in the case that small phase
and flux calibration errors are still present in the data. The
resulting map at 1.4 GHz is shown in Fig.1. All maps will be presented
in more detail in forthcoming publications. The 1.4--GHz map is used
as a representation of the surface brightness distribution against
which we observe the HI distribution in the lens galaxy. We use it to
calculate the HI absorption line. We note that absorption really
occurs at $1.42/(1+z_l)=0.75$~GHz, for which we do not have a radio
map. However, the difference between the 1.4 and 1.7 GHz maps are
small and we expect the radio map at 0.75 GHz to be quite similar to
that at 1.4 GHz. Using the 1.7 and 1.4 GHz radio maps to construct an
estimated 0.75--GHz radio map is difficult given the different UV
sampling. We therefore opt to solely use the 1.4--GHz map.

The multi-frequency frontends of the WSRT were used to obtain a high
signal-to-noise profile of the HI absorption line at a frequency of
753 MHz. Compared to the previous observations of Chengalur et
al. (1999) the data have superior spectral resolution and wider
bandwidth as well as a better spectral baseline.  In total we obtained
5.5~hrs of data on PKS1830--211 with 12 MFFE's on October 25, 1998. A
backend bandwidth of 2.5~MHZ was used with 256 channels, giving
3.9~\kms\ per channel. The data were Hanning smoothed to give a
resolution at halfwidth of 7.9~\kms. The resulting normalized line
profile is shown in Fig.2. Errors are estimated from the rms scatter
between channels at the baseline level well away from the line
profile.

\section{Kinematic Model}

We model the HI absorption line, assuming a thin axisymmetric HI disk
with an optical depth $\tau(r)=\tau_0\times (r/r_0)^\gamma$. The disk
has a kinematic center $(x_k,y_k)$ on the sky, an inclination $i$ and
position angle $\theta$ (measured north to east on the sky). The HI
gas has a constant circular velocity $v_c$, as expected for an
isothermal mass distribution of the lens galaxy and a systemic
velocity $v_s$. The latter allows for a small velocity difference with
respect to the redshift determined from molecular lines. For each
surface brightness element $\Sigma_v(\vec{x})$ of the lensed radio
source [$\vec{x}=(x,y)$ are coordinates on the sky], we then
calculate~(i) the radial component $v_r(\vec{x})$ of the velocity
vector of the HI gas, where the line-of-sight pierces the disk, and
(ii) the optical depth $\tau(\vec{x})$ for the sight line normal to
the disk, which is then multiplied by $\cos(i)^{-1}$ to include the
effect of inclination.  We assume that the HI absorption line is
isotropically (Gaussian) broadened by $\sigma_{\rm HI}$ (=FWHM/2.35),
constant as a function of radius.

We re-normalize the flux-density of the source, $\int
\Sigma_v(\vec{x}) {\rm d}^2\vec{x}$, to unity.  The observed line
profile becomes a convolution of a Gaussian velocity profile with the
unbroadened line-profile\footnote{This is correct only if the
$\sigma_{\rm HI}$ is not a function of $\vec{x}$ and note also that
the Gaussian is not normalized.}, i.e.:
\begin{equation}
      I(v)=\left\{{e^{v^2/2 \sigma_{\rm HI}^2}}\right\} \star
      \left\{\int_{v\equiv v_r(\vec{x})}
      \Sigma_v(\vec{x})\,\left[1-\tau(\vec{x})\right] ~{\rm d}\vec{x}
      \right\}
\end{equation}
which is unity in the case of no absorption. Images of the radio
sources -- even though they have a continuous mapping and surface
brightness distribution -- are always affected by the finite beam
size. Hence, instead of using the MERLIN image of PKS1830$-$211 to
model $\Sigma_v(\vec{x})$, we instead use the clean
components\footnote{Clean components are delta functions with finite
flux that, convolved with a Gaussian beam, reproduce the radio map as
presented in Fig.1. In total our model of the source brightness
distribution includes 350 clean components.}. In this case the
integral in Eq.1 is simply replaced by a sum over the clean
components. The resulting absorption line profile should therefore
more closely resemble the underlying line profile\footnote{The
line-width $\sigma_{\rm HI}$ might increase slightly because of
unaccounted for minor velocity gradients between clean components.}
for the deconvolved radio image of PKS1830$-$211. The free parameters
in our kinematic model are $\{i,\theta,v_s,\sigma_{\rm
HI},\tau_0,\gamma\}$. We furthermore set $v_c=\sqrt{2}\times
\sigma_{\rm DM} = 266$\,\kms, which follows from the isothermal lens
mass model (i.e. a dark-matter dispersion of $\sigma_{\rm
DM}$=188\,\kms; in agreement with Winn et al. 2002), $r_0=0\farcs5$
(arbitrary value but roughly the Einstein radius), and the kinematic
center is set to the lens centers inferred from the different authors
discussed in Sect. 1. We register the optical and radio data,
associating the bright radio cores with the optical quasar images.

\section{Results}

Constraints on the kinematic model are the following: (i) the WSRT HI
line-profile with 238 channels with 3.9~\kms\ width
each, with an average 1--$\sigma$ measurement error of 0.00147 (Fig.2;
three channels are removed due to band-pass edge uncertainties), and
(ii) the velocity difference between components A and B of
147$(\pm$1)\,\kms\ (Wiklind \& Combes 1998). We minimize $\chi^2$ -- as
commonly defined -- using a Downhill Simplex method with simulated
annealing (Press et al. 1992), varying all six free parameters
simultaneously.

We only explore models where the kinematic centers have been fixed at
positions (notably often mutually exclusive) as previously published
in the literature. Given degeneracies in the kinematic model, the
current spatially unresolved line-profile and potential other issues
(see below), we feel that a more sophisticated kinematic model is at
this point not warranted.

\begin{table*}
\centering
\begin{tabular}{ccccccccl}
\hline
\hline
     $x_k$ ($''$) & $y_k$ ($''$) & $i$ ($^\circ$) & $\theta$ ($^\circ$) &  $\sigma_{\rm HI}$ (km\,s$^{-1}$) & $\gamma$ & $\tau_0$ & $\chi^2$ & Reference\\	
\hline
     $+$0.532  &  $-$0.420  &   30  &   $-$15     &  40   &   0.76   & 0.076  & 443 &  Nair et al. (1993)\\
     $+$0.519  &  $-$0.511  &   17  &   $+$23     &  48   &   0.65   & 0.078  & 446 &  Courbin et al. (2002), G\\     
     $+$0.500  &  $-$0.450  &   32  &   $-$15     &  39   &   0.66   & 0.079  & 443 &  Lehar et al. (2000)\\
     $+$0.350  &  $-$0.510  &   17  &   $+$28     &  48   &   2.14   & 0.065  & 446 &  Kochanek \& Narayan (1992)\\
     $+$0.328  &  $-$0.486  &   20  &   $+$11     &  47   &   3.62   & 0.057  & 449 &  Winn et al. (2002)\\  
     $+$0.285  &  $-$0.722  &   21  &   $+$34     &  48   &   1.73   & 0.049  & 449 &  Courbin et al. (2002), Sp\\
\hline
\hline
\end{tabular}\\~\\
Table 1 --- Best-fits parameters of the kinematic models (see Sect.\,3--4, also for errors).
\end{table*}

The results are listed in Table~1. A number of conclusions can be
drawn:

\medskip\noindent {\bf (1)} The inclination of the HI disk,
$i=(17-32)^\circ \pm 2^\circ$ (1--$\sigma$), is robust and nearly
independent from the assumed kinematic center of the lens galaxy. This
confirms that the lens galaxy is most likely a close-to face-on spiral
galaxy, as previously proposed (e.g. Winn et al. 2002).

\medskip\noindent {\bf (2)} The position angle of the disk varies
between $\theta = +34^\circ$ and $-15^\circ$.  This result is clearly
less robust than the inclination angle, due to degeneracies with the
inclination and kinematic center.

\medskip\noindent {\bf (3)} The line-width, $\sigma_{\rm
HI}=(39-48)\pm 1$~\kms, is also robust. The value appears high
compared with local spiral galaxies and might be the result of
increased star formation, stirring up the HI gas to higher turbulent
velocities. The line-width of molecular gas is also relatively high,
$\sim$30~\kms (Wiklind \& Combes 1996). Even though lower than what we
infer, such difference can be expected between colder molecular gas,
presumably in clouds, and HI in the atomic (versus molecular)
phase. However, the use of clean components to represent the, in
reality, smooth Einstein Ring could somewhat broaden the fitted
line-width because they undersample the velocity field (i.e.~the field
in between clean components will shows a velocity gradient that is not
sampled). The spectral resolution is negligible.

\medskip\noindent {\bf (4)} The optical depth varies considerably as a
function of radius for some of the models, with typical 1--$\sigma$
statistical errors on $\tau_0$ of $\pm$0.01. In particular the models
with the lens centers from Winn et al. (2002), Kochanek \& Narayan
(1992) and Courbin et al. (2002; only for model Sp) seem to prefer
steeply {\sl rising} optical depths as a function of radius
(i.e. $\gamma \gg 0$). The other models also prefer rising optical
depths, but more moderately. All models give a considerable HI
depletion toward the lens center, as is also seen in some local
galaxies, which could be the result of strong past and ongoing star
formation (see also --3-- above). This depletion can already been
inferred from the fact that, even though image B is closer the
inferred lens centers, it experiences less HI absorption (Fig.2).

\medskip\noindent {\bf (5)} Plotting $\tau(r)$, we find that all
models give $\tau\approx 0.085$ at $r=5.0\,h^{-1}_{70}$~kpc along the
disk, except for the lens center (Sp) from Courbin et al. (2002) for
which 0.06 is found.  However, from a lensing perspective it is
unlikely that the latter position is that of the dominant lens galaxy.
The HI column density is $N_{\rm HI}= 4.6\times 10^{21}\, (T_{\rm
s}/100~{\rm K}) \,(\sigma_{\rm HI}/{{\rm km}\,{\rm
s}^{-1}})\,\tau$~cm$^{-2}$ (given our definition of the line profile
in Eq.\,1). Assuming a ``canonical'' value of the unknown spin
temperature of $T_{\rm s}=100$~K, as in Chengalur et al. (1999),
$\tau\approx 0.085$ at a radius of 5\,$h^{-1}_{70}$~kpc in the disk
and $\sigma_{\rm HI}\approx 44$~\kms, we find $N_{\rm HI}\approx
2\times 10^{21}$~cm$^{-2}$, similar to what was already found by
Chengalur et al. and also similar to that of local spiral galaxies and
damped Ly$\alpha$ systems.

\begin{inlinefigure}
\vspace{0.3cm}
\begin{center}
\resizebox{\textwidth}{!}{\includegraphics{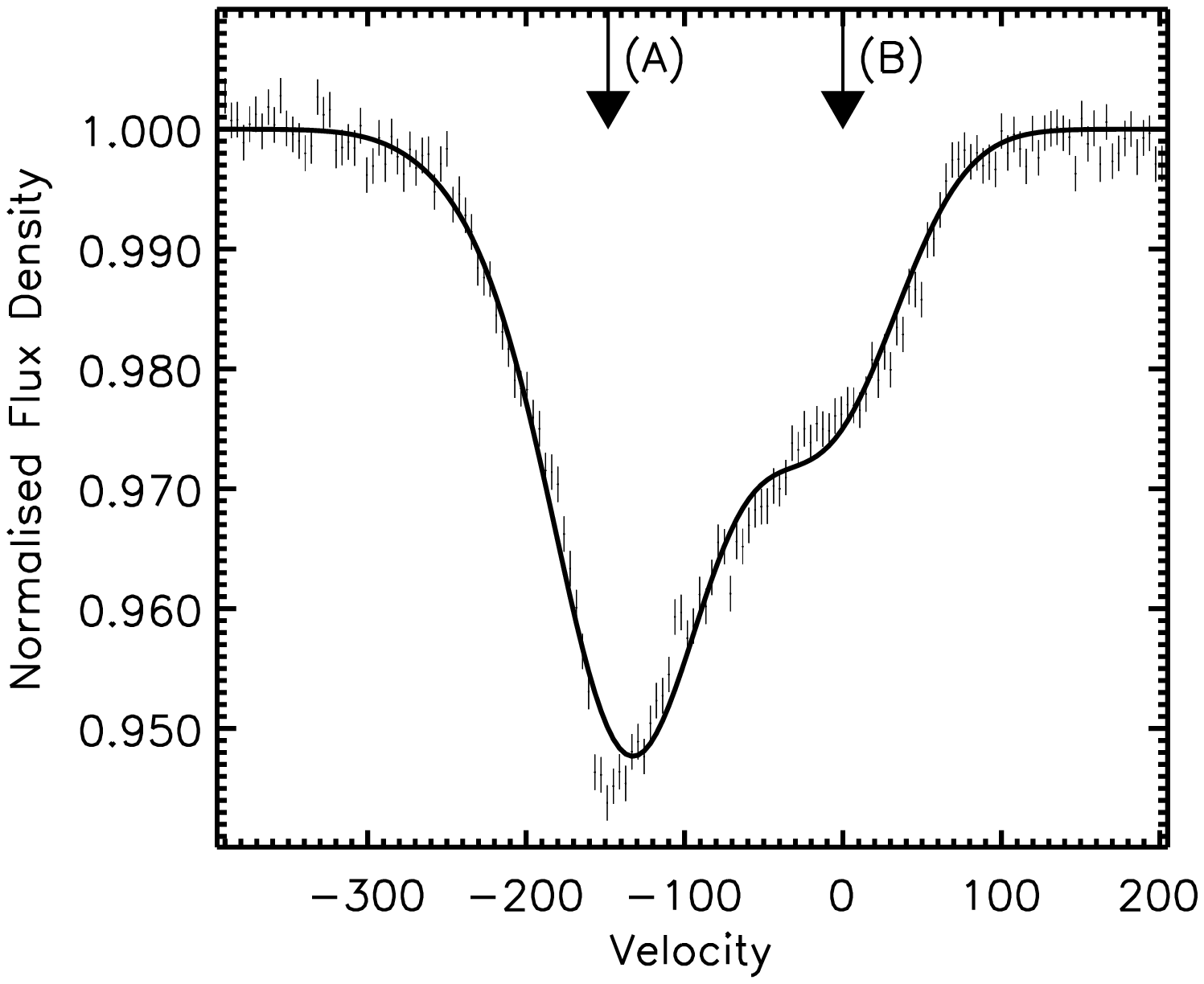}}
\end{center}
\figcaption{The WSRT HI--line profile overplotted with the best-fit
kinematic model from Fig.1. Errors are 0.00147 in normalised units.
The velocity is in units of \kms. A and B indicate the velocity
positions of the two lensed cores of source.}
\end{inlinefigure}

\medskip
We note that the range of parameter values are typically larger than
their 1--$\sigma$ errors, since the former reflects the change as
function of chosen kinematic center.

The results have to be seen in light of some additional cautionary
notes: First, we see that the $\chi^2$ values are $\sim$2~times the
DOF=229. This is mostly due to a mismatch between the model and the
observed HI line around A and B (see Fig.2). The lensed core of the
source, however, is known to be time-variable (Lovell et
al. 1998). Since, the MERLIN and WSRT data set were not obtained at
exactly the same epoch (5~months in between), this might explain the
mismatch. WSRT data from different epochs (not presented here) indeed
show some minor differences between the observed HI lines.

Second, our models are azimuthally symmetric and assume an HI covering
factor of 100\%. If the HI column density (i.e. optical depth) is a
function of angle along the disk, it could change the inferred
kinematic results. Even though we believe the effect to be small, it
can not fully be excluded. A spatially resolved HI absorption line
profile would help to eliminate this possibility.

Third, we have assumed the 1.4--GHz maps to be a reasonable
representation of the 0.75--GHz continuum structure of the lensed
source, as discussed previously. If the spectral index is a strong
function along the lensed source, it might also affect the results.

Fourth, we have assumed that the rotation curve of the spiral galaxy
is flat. This is consistent with observations of local spiral galaxies
with similar rotation velocities, but might not be a correct
assumption for a galaxy at $z=0.89$. Again, a spatially resolved HI
absorption line profile would be able to test this assumption.

\section{Discussion \& Conclusions}

We have modeled the neutral hydrogen (HI) distribution and kinematics
in the spiral galaxy of the lens system PKS 1830$-$211 at $z$=0.89,
using a MERLIN 1.4--GHz continuum map of the Einstein ring and an WSRT
HI absorption line. The main numerical results are summarized in
Sect.~4.

In general, we find that the HI disk is seen close-to face-on, as
previously suggested (e.g. Winn et al. 2002). We find evidence for HI
depletion toward the galaxy center and a column density of $N_{\rm
HI}\approx 2\times 10^{21}$~cm$^{-2}$ at 5\,$h^{-1}_{70}$~kpc,
comparable to those of low redshift spiral galaxies. Our data are
insufficient to discriminate between the different lens centers as
published by different authors and we have to await new HST--ACS
observations.

We emphasize some cautionary notes, listed in Sect.4, regarding our
results.  Even so, we believe that our analysis is a good first
attempt to constrain the kinematics of neutral hydrogen in a spiral
galaxy at relatively high redshift, within the current observational
limits, and should be regarded as such.

Our analysis shows that the combination of HI observations with
gravitational lensing -- in particular radio Einstein rings --
potentially provide a strong tool to study spiral galaxies to high
redshifts.  A similar program to combine lensing and stellar
kinematics of E/S0 galaxies out to $z\sim1$ is currently underway
(i.e. the Lenses Structure \& Dynamics (LSD) Survey; e.g. Koopmans \&
Treu 2002, 2003; Treu \& Koopmans 2002).

We anticipate that {\sl spatially resolved} HI absorption line
profiles and radio continuum images at the redshifted HI-line
frequency can be obtained of many similar systems in the future with
e.g. ALMA, LOFAR and SKA. This would resolve many of the issues
outlined in the paper and provide us with an unprecedented probe into
the internal structure of spiral galaxies at high redshifts. By
probing spiral lenses over a range of redshifts, their evolution can
be studied (e.g. Koopmans \& de Bruyn 1999), providing a complementary
tool to study spiral galaxies between those at low redshifts using HI
emission and damped Ly-$\alpha$ systems at very high redshifts.

We note, however, that none of the other studies can provide {\sl two
independent} constraints, i.e. from lensing and kinematics, on the
same mass distribution of the spiral galaxy. Their combination can
break degeneracies inherent to each technique separately. Our results,
admittedly based on a single system, show that very interesting
constraints can be obtained for spiral galaxies at high redshifts.

\acknowledgments We like to thank Tom Muxlow for doing the initial
calibration of the MERLIN data. We also thank Eric Agol, Ron Allen,
Mike Fall and Mario Livio for comments and discussions. LVEK
acknowledges the support from an STScI Fellowship grant. MERLIN is a
National Facility operated by the University of Manchester at Jodrell
Bank Observatory on behalf of PPARC. The Westerbork Synthesis Radio
Telescope (WSRT) is operated by the Netherlands Fouondation for
Research in Astronomy (ASTRON) with the financial support from the
Netherlands Organisation for Scientific Research (NWO).

\clearpage

\end{document}